\documentclass[a4paper]{llncs}
\usepackage{amsmath}
\usepackage{amssymb}
\usepackage{latexsym}
\usepackage{stmaryrd}
\usepackage{pgf}
\usepackage{tikz}
\usetikzlibrary{automata}
\usepackage{graphicx}
\usepackage{subfigure}
\usepackage{subfig}

\usepackage{url}

\newcommand{\keywords}[1]{\par\addvspace\baselineskip
\noindent\keywordname\enspace\ignorespaces#1}

\begin{document}
\mainmatter

\title{A Checklist for the Evaluation of Pedestrian Simulation Software Functionalities}

\titlerunning{A Checklist for the Evaluation of Pedestrian Simulation Software Functionalities}

%
%
\author{Mizar Luca Federici\inst{1}, Lorenza Manenti\inst{1,2}, Sara Manzoni\inst{1,2}}
\authorrunning{M. L. Federici, L. Manenti, S. Manzoni}

\institute{
CROWDYXITY s.r.l. - Crowd Dynamics and Complexity\\
Via Rombon 41\\
20134, Milano, Italy\\
\url{m.federici@crowdyxity.com}
\and
Department of Informatics, Systems and Communication\\
University of Milano-Bicocca\\
Viale Sarca 336\\
20126, Milano, Italy\\
\url{{manenti, manzoni}@disco.unimib.it}
}

\maketitle

\begin{abstract}

The employment of micro-simulation (agent-based) tools in the phase of design of public and private spaces and facilities and for the definition of transport schemes that impact on pedestrian flows, thanks to their achieved accuracy and predictive capacity, has become a consolidated practice. These instruments provide support to the organization of spaces, services and facilities and to the definition of management procedures for normal and emergency situations. The employment of these tools is effective for various but not for all the contexts, nevertheless new features and functions are under constant development and new products are often launched on the market. Therefore, there is a higher necessity of a standard criteria both for the evaluation of the kinds of function that these software provide, at use of practitioners and end-users, and for the definition of software requirements as a reference for the developers that aim at being competitive on this market.

On the basis of our experience as pedestrian modellers and as researchers in the crowd modelling area, we designed a comprehensive and detailed ready-to-use checklist for the quantitative evaluation of Pedestrian Simulation Software functionalities that aims at capturing all the aspects that we claim that are useful to undertake a professional study. These functions in our opinion are necessary to provide accurate results in the planning of new facilities or schemes that involve pedestrian activities. With this work we propose a set of criteria of evaluation for these products also to encourage a debate for the definition of objective standards for pedestrian simulation software certification.
\end{abstract}

\keywords pedestrian simulation, agent-base models, evaluation, crowd simulation

\section{Introduction}

Simulation studies of crowds and pedestrian dynamics are nowadays more and more required in the phases of assessment and design of public and private spaces and facilities and they constitute valid tools for the support of the work of Architects, Engineers, Urban Planners, Decision makers, Security managers and so on. The use of simulation instruments successfully shows advantages in heterogeneous contexts such as transit stations (underground, railway, bus terminal), airports, sport venues (stadium, arenas, urban sport events), retails (shopping centers, supermarkets), private and public buildings (schools, prisons), entertainment (theaters, cinemas, concerts) and large events. Micro-Simulation tools have therefore become sufficiently reliable to predict with accuracy, in specific contexts of application (and if informed with a reliable demand), the characteristics of pedestrian dynamics and flows and to highlight in advance possible criticalities. They also allow to: dimension the number of pedestrian facilities or operative components, like gates or Vertical Circulation Elements (VCE) for normal and emergency operations; predict egress and evacuation times; assess the performance of different functional areas of the environment; evaluate operative procedures; compare the efficiency of different layouts or configurations of the environment; study worst-case/what-if scenarios and disruptions that are not directly observable in reality; support the definition of strategies of crowd management and decision making; optimize the positioning of specific services and of sign-posting; assess the accessibility and movement of People with Restricted Mobility (PRMs); perform social cost analysis etc. 



\begin{figure}[h]
	\centering
	\includegraphics[width=\columnwidth]{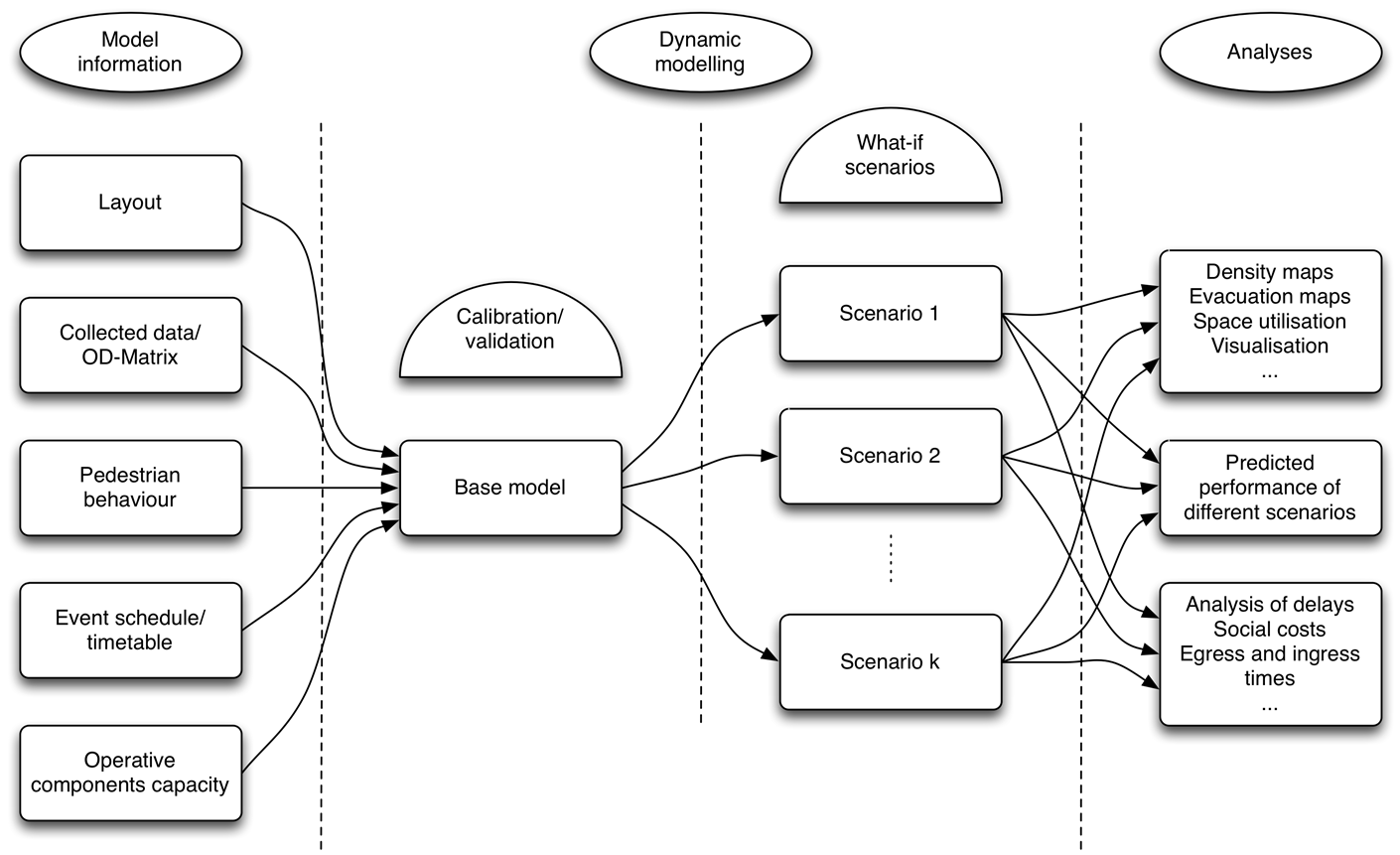}
	\caption{The simulation process}
	\label{fig:flusso}
\end{figure}

The scheme in Fig. \ref{fig:flusso} shows the typical workflow of the simulation process. The first step is the information of the model. The great part of these platforms allows to import the layout of the environment in a CAD drawing form and to import Origin-Destination matrices (OD matrices) of pedestrian flows combined with an event schedule. These data inform the model with the observed or predicted demand. The demand is usually prepared on the basis of a data collection campaign performed onsite (i.e. people headcounts in a station). If the simulated environment does not exist yet, simulation tools can still be used in the design process for the comparison of predicted performances of alternative options (layout, services, demand etc.). In this case the demand has to be shaped by techniques of demand forecasting.

Then, in the phase of dynamic modelling, the base model has to be calibrated and validated. Only then variations of the base model (i.e. what-if scenarios that might include alternative design, time schedule variations or evacuation scenarios) can be considered reliable. 

In the phase of analysis, simulation analysis tools allow extracting quantitative and qualitative data from simulation runs. Pedestrian  dynamics can be visualized in 2D or 3D and outputs can be extracted in a numerical form or analyzed in the form of charts or maps of various kinds. Data that can usually be extracted from these software are related to density, occupancy times, transfer times, delays and so on.

The visualization of pedestrian and mobility dynamics provided by some pedestrian simulation software can be extremely advanced but it has to be underlined that the accuracy of the reproduced dynamics, and the consequent prediction of the performance of the environment and of the procedures and operations under study, relies on the accurate information of the simulation model, on the exact reproduction of the layout under study and on a proper calibration and validation of the model.

At our knowledge it does not exist a general evaluation criteria of these platforms that reflects the simulation process performed according to the simulation best practices (see \cite{nikoukaran1998criteria} for a more general work on the evaluation of simulation software). We believe that the ready-to-use checklist that we propose might provide a first draft of a standard criteria that would constitute an added value both for the practitioners, in choosing the appropriate platform to accomplish their needs, and to developers of new commercial software that aims at being competitive on the market.

In particular we think that the proposed checklist would be useful to evaluate the completeness of a proposed software against a comprehensive list of possible functions. These functions have been determined on the basis of what we perceive, under the practitioner's point of view, as the tools that are fundamental to undertake an accurate pedestrian modelling study.


Our impression is that in spite of the many offers of pedestrian simulation software present on the market, none of these refers to a specific standard of what a pedestrian simulation software should do. 

Our intention is therefore to propose a common soil of comparison between the different software.  Although what we propose is a quantitative checklist that summarizes the functions that a pedestrian simulation software might have, without offering a method to judge also qualitatively each specific function, we believe that the quality of the studies undertaken with these tools is definitely highly dependent on the modelling and analysis options made available in each software, in addition to the modeller's skills. A lack of expressive capacity for the definition of the environment, pedestrian dynamics or scheduled events and poor analysis possibilities, in our opinion, inevitably ends up in an excessive approximation of reality, in the simulation, that would reflect also on the accuracy and quality of the final results of the study.

Although we have applied our criteria of evaluation to a few software and we found it useful and effective, also if improvable, here we will not refer to any specific software, differently from works like \cite{castle2011comparison} \cite{xiao2005methodology} \cite{haron2012software} that compare the use of specific competitor platforms on the same scenario. 

Some high-level guidelines user-oriented for the choice of evacuation models/software can be also found in \cite{kuligowski2005user}. In \cite{castle2007guidelines} the author instead proposes some key questions a final user should answer to choose accurately an evacuation software after introducing in details different formal models that the platforms can embedded. See also \cite{AlexanderssonJohansson:Thesis2013} for similar criteria of comparison declined on two specific software.

On the basis of our experience as users (in terms of pedestrian modelers) of commercial platforms and as researchers in the crowd modelling area, we designed a detailed checklist for the evaluation of Pedestrian Simulation Software that aims at capturing all the aspects that we believe as necessary to undertake a professional study and, therefore, to provide accurate results in the planning of new facilities or transport schemes that have impact on pedestrian flow. With this work we propose a set of criteria of evaluation for these products that also aims at starting a debate for the definition of objective standards for pedestrian simulation software certification.

This work is organized as follows: Section 2 will introduce at high-level the aspects that we believe are important for the evaluation of the characteristics and functionalities of a pedestrian simulation software; Section 3 will list in details, for each of the considered aspects, the single functionalities that are for us relevant and it will offer a brief description of each aspect with the purpose of making this section accessible also to final users that are not experts of the topic; Section 4 will briefly describe the proposed quantitative evaluation criteria; paper ends with final remarks and future works in Section 5.

\section{Requirements: High-level Categorization}
\label{sec:overview}

In this Section we introduce in a form of high-level categorization organized in three subsets the features that we think should be supported by different pedestrian simulation platforms. Every subset collects all the features which support specific aspects of the process of the definition of a pedestrian simulation. 		

\begin{figure}[!hb]
	\centering
	\includegraphics[width=\columnwidth]{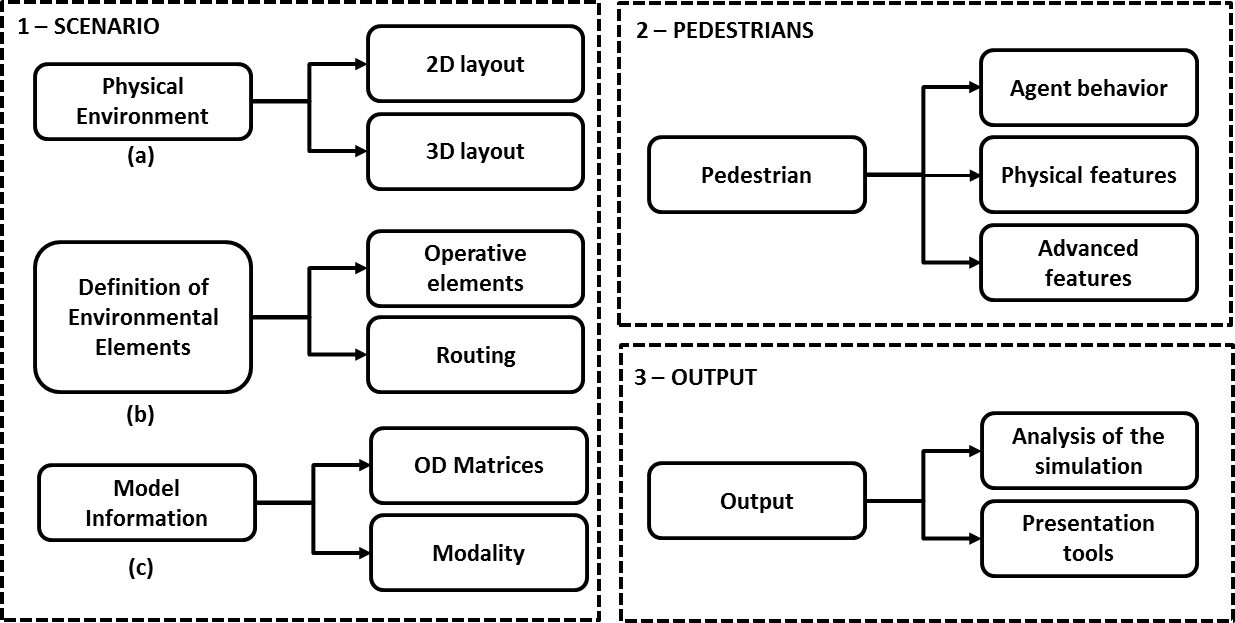}
	\caption{The scheme above summarizes the high-level categorization}
	\label{fig:overview}
\end{figure}

The three subsets are identified as follows (see Fig. \ref{fig:overview}):

\begin{enumerate}
\item SCENARIO (subset 1): functions supporting the definition of the simulated \emph{scenario}. In this subset we refer to all the aspects dedicated to the preparation of the scenario where the dynamic simulation will take place. The scenario definition implies the possibility of manipulation of the physical environment (e.g. by means of CAD files that can be imported and manipulated to obtain the basic environment of the simulation) (subset 1\emph{(a)}), the characterization of its operative elements (stairs, escalators, lifts, ticket windows and so on) and of the routing elements that would guide pedestrians through the environment and that would influence their decisions (subset 1\emph{(b)}). The scenario definition also includes the possibility to inform the model by means of OD-matrices and the possibility to vary the mode of the simulation (subset 1\emph{(c)});

\item PEDESTRIANS (subset 2): functions supporting the possibility to model different types of pedestrians by varying their physical characteristics as speed, size, encumbrance, and also their behavioral components such as decision making processes. Some advanced features might include the cohesion of a group, a partial \emph{a-priori} knowledge of the environment and the possibility to define emotional states;

\item OUTPUT (subset 3): functions supporting the extraction of simulation \emph{output} to perform analysis and to export and visualize data related to density, delay times, distances, pedestrian counting and so on. A set of tools for the management and the presentation of outputs in the form of maps and charts and that includes the possibility of performing some data filtering should be provided by the platform.
\end{enumerate}

In the next Section we describe each of the above listed subsets in details.

\section{Detailed Checklist for Pedestrian Simulation Software Functionalities}
\label{sec:checklist}

In this section we describe the details of our checklist that  we have conceptually structured as follows:

\begin{itemize}
\item 2D physical environment definition;
\item 3D environment definition; 
\item advanced environmental features; 
\item modelling and routing objects;
\item OD-matrix input and manipulation; 
\item evacuation studies;
\item pedestrian characteristics and behaviour; 
\item analysis of simulation outputs; 
\item presentation tools;
\item vehicle and pedestrian interaction; 
\item robustness; 
\item validation.
\end{itemize}

A description is provided for all the functionalities that we think should be considered (see Fig. \ref{fig:details} for a graphical representation of the checklist). Between these all the voices signed with * are the ones that constitute, in our opinion, mandatory requirements. 

\subsection{SCENARIO: Requirements (subset 1)}

\subsubsection{2D Physical Environment definition (subset 1\emph{(a)})}

The acquisition of the 2D layout of the environment constitutes the very first step of a project and it is normally provided in the form of an CAD drawing or in images that depict the layout at a specific scale. Moreover basic simulation dynamics take place in 2D, although they might be then displayed (for some of these software) directly in 3D. For this reason we think that the possibility to import and define with accuracy the 2D environment, and, therefore, to reproduce faithfully the layout where the simulation has to take place is one of the features that has to be considered as fundamental. Also, due to the fact that physical space can be, in relation to the adopted approach, discrete or continous and this might imply differences in the modelling procedures, we also would like to keep track of this characteristic.
Our checklist for the definition of the 2D environment includes the following controls:

\begin{itemize}
\item space representation:

	\begin{itemize}
	\item continuos: space is continous;
	\item discrete (grid/cells): space is discrete, each agent occupies a cell of a grid, or more cells;
	\item grid size is variable: in case of discrete space representation the dimension of the cells is customizable;
	\end{itemize}

\item 2D CAD import:
	\begin{itemize}
	\item 2D CAD import as reference only*: CAD drawing in .dwg/.dxf or other format can be imported in scale but just as reference, the environment is built using the software pre-defined objects;
	\item 2D CAD import as obstacle: CAD drawing in .dwg/.dxf or other format can be imported in scale and CAD lines constitute \lq\lq physical\rq\rq\ obstacles (or cell obstructions in case of discrete space representation) in the simulation;

	\end{itemize}
	
\item 2D CAD manipulation: the software allows basic CAD functions to modify the imported layout. CAD drawing can be cut, copied, pasted, rotated, translated and scaled inside the simulation platform. New CAD lines and poly lines (including circles, squares and other kinds of polygons) can be drawn. CAD colour can be changed;

\begin{figure}[!h]
	\centering
	\includegraphics[width=0.99\columnwidth]{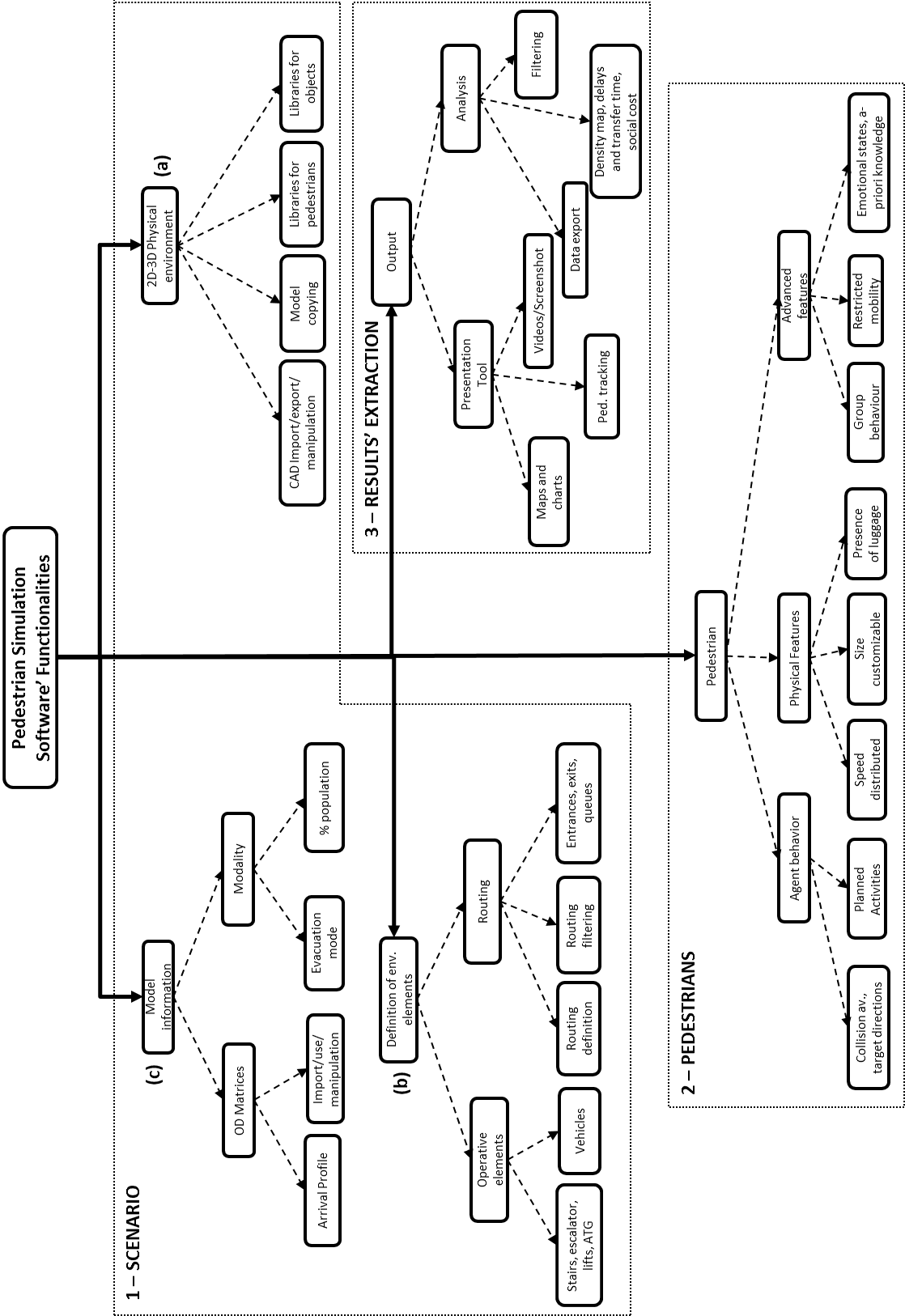}
	\caption{Detailed view of pedestrian simulation software functionalities}
	\label{fig:details}
\end{figure}

\newpage
\item 2D CAD layers:
	\begin{itemize}
	\item CAD drawing can be organized in layers: original CAD layer can be maintained in the import process into the pedestrian simulation software. New CAD layers can be created and objects can be moved across layers inside the platform;
	\item simulation layers: a layer can be selected or not to be used as obstacle in simulation; 
	\end{itemize}
	
	\item 2D CAD export: the software allows to export the CAD drawing into .dwg/.dxf or other format once it has been imported and after it has been modified inside the simulation software;

\item 2D CAD copy and paste across models: CAD objects (and layers) can be copied across different files;

\item 2D CAD picture (.tiff/.jpg/.bmp or other format) import: to be used as reference or for presentation;

\item measuring in the model*: possibility to measure distances and angles respectively  with a \lq\lq ruler\rq\rq\ and a protractor function inside the platform;

\end{itemize}

CAD import and manipulation inside the platform, and the possibility to consider CAD lines as obstacles (or cells inhibitions in case of discretized space) for pedestrians is in our opinion strongly recommended. In particular, the possibility to modify the CAD drawing  inside the platform is extremely useful as it limits the use of external AutoCAD software. Revisions (sometimes also minor) to the layout are often required. We would like to add that, although the list above is mainly referred to CAD import and manipulation, in case the pedestrian simulation software implies the use of its own obstacle objects only, to shape obstacles, we consider this solution acceptable but we would recommend to add the possibility to copy, paste, rotate, scale, export, modify as well these objects once they have been drawn. In particular very useful and time-saving is the possibility to copy across different models.

\subsubsection{3D Environment definition (subset1\emph{(a)})}

3D visualization of simulation dynamics can be particularly effective especially for presentation purposes and, in the case of complex structures (i.e. multi-floor buildings) 3D rendering helps the understanding of simulation dynamics, especially for non technicians.
Although we do not consider 3D visualization essential for pedestrians simulation studies, we list below the characteristics that need to be considered in case the software platform allows the opportunity of this form of modelling/visualization:

\begin{itemize}

\item 3D environment representation:
	\begin{itemize}
	\item supported in the platform: the platform provides tools to build the 3D model of the simulation environment;
	\item external platform needed: 3D modelling is possible but only if the 3D environment is built in another platform or an additional module of the software is needed (this is the case that requires the construction of the 3D scenario in an external platform - i.e. 3D studio Max, Blender, SketchUp Pro - and the use of an additional module to combine the 2D simulated data and the 3D scenario also to animate the pedestrians);
	\end{itemize}
	
\item 3D pedestrians: pedestrians can be modelled in 3D and can be selected from a library of predefined types;
\item 3D import: the platform allows to import .dwg/.dxf/.3ds/.dae or other formats of 3D CAD drawing;
	
\item 3D manipulation: 3D CAD objects can be built and manipulated (copied, pasted, rotated, scaled and so on) inside the platform and also across models;
	
\item 3D layers: 3D objects can be organized in layers; 

\item 3D export: manipulated CAD can be exported from the simulation into .dwg/.dxf/.3ds/.dae or other formats;

\item 3D object animation:
	\begin{itemize}
	\item modelled in the platform: objects (i.e. vehicles, escalators, lifts) can be or are animated directly in the simulation platform;
	\item importable from other platforms: 3D animations (i.e. embedded in 3ds files) can be imported and matched with the simulation environment;
	\end{itemize}

\item 3D animated pedestrians:
	\begin{itemize} 
	\item modelled in the platform: walking pedestrians are modelled in the environment;
	\item importable from external sources: importable from pre-defined libraries and super-imposable to 2D simulation;
	\end{itemize}

\end{itemize}

\subsubsection{Advanced Environmental Features (subset 1\emph{(b)})}

Besides modelling objects of general use that can be be combined to create specific functional areas or operative components of the environment together with complex mechanism (i.e. boarding and alighting) and that might be available in the software, the availability of pre-defined customizable objects, like stairs, escalators or lifts (in some platforms also counters, automatic gates and so on) could be extremely helpful to modellers. We list below some of the features that we consider useful:

\begin{itemize}

\item built-in objects (VCE):
	\begin{itemize}
	\item stairs: stair objects are pre-defined but customizable in width and height and the modelling of handrails is allowed. Possibility to shape landings and multiple flights of stairs;
	\item escalators: escalator objects are customizable in width, height and persons per minute (ppm) capacity;
	\item lifts: lift objects are customizable by capacity, fixed cycle or also by an algorithm that manages the priority in case of concurrent calling from different floors (e.g. some software provide an advanced lift modeling while it is notorious that in other simulation platforms the modelling of multi-floor building might require very complicate and time spending modelling artifacts to represent lifts);
	\end{itemize}

\item vehicles:
	\begin{itemize}
	\item train: train objects can be selected from a library and are customizable by number of carriages, carriage capacity, doors number and position. Passengers can interact with the vehicles by alighting and boarding;
	\item car: car objects can be chosen from a library and can be animated;
	\item bus: bus objects chosen from a library can be boarded and alighted;
	\item plane: plane objects can be chosen from a library and can be boarded and alighted;
	\item ship: ship objects can be chosen from a library can be boarded and alighted;
	\end{itemize}
	
	\item context specific objects: pre-defined libraries of objects (e.g. airport related objects as luggage belt, check-in and check-out areas, ticket counters) that are not easily modelled with general purpose objects.
	
\end{itemize}

\subsubsection{Modelling and Routing Objects (subset 1\emph{(b)})}

Modelling and routing objects' characteristics and solutions are fundamental to represent the dynamics in the simulation, especially to give account of complex behaviors (i.e. changing of final target at the fulfillment of some conditions; queuing; boarding and alighting in correspondence of specific events). This section aims at specifying what we believe are the main modelling and routing functions that need to be implemented in the software and that we consider necessary for capturing the major part of possible behaviors and dynamics\footnote{we have assumed that the great part of the decision making process is delegated to the environment. Probably for computational reasons it seems that this approach is the most commonly adopted in these kinds of software. Nevertheless we believe that also the opposite paradigm that defines all the decision making and route choice selection based on a representation of the environment hosted "`inside"' each single agent is absolutely valid}. 

\begin{itemize}
\item Modelling objects should be customizable to cover the following functions:
	\begin{itemize}
	\item input areas*: objects that allow to input specific types of pedestrians at a specific rate or following a specific timetable;
	\item exit areas*: objects that when reached make agent disappear from the simulation;
	\item target areas* (markers) or similar: objects that can constitute intermediate target in the route of a pedestrian and that, once reached, assign a new target to the pedestrian (with the possibility to split by percentage or according to other criteria);
	\item waiting areas* or similar: areas that can be assigned as a target to the pedestrians and that once reached distribute pedestrians inside their perimeter following a customizable criteria (i.e. to model platform);
	\item zones that change pedestrian behavior: zones or areas that when crossed (if they are placed in the middle of some paths) influence the behavior of pedestrians, for example slowing them down, speeding them up, \lq\lq pushing\rq\rq\ them in one direction against shortest path or obliging them to a circular movement of customizable radius. These zones or objects represent a bias to the normal behavior of pedestrians and their action could be selective on specific pedestrian's types;
	\item \lq\lq delay\rq\rq\ or \lq\lq stop\rq\rq\ areas* or similar: objects that can be assigned as target to the agents and that once reached hold the pedestrian for a specific time. These objects can be used to delay passengers that go through an Automatic Ticket Gate (ATG, fixed delay) or that are purchasing a ticket in front of a Ticket Machine (TM, variable delay);
	\item modifiers of target or of pedestrian kind: objects that if crossed influence the behavior of the pedestrians by assigning them a different target or changing their type. These objects could be set to work following a specific timetable, always, or only if specific conditions are fulfilled (i.e. if an area is congested arriving pedestrians are re-routed towards a less congested path). For this last option these objects have to be coupled with analysis that monitors the condition present in specific portion of simulated environment. These objects might also affect only specific pedestrians (filtering by type of pedestrians);
	\item queuing areas*: areas that organize pedestrians in ordered queue and that manage the priority;
	\end{itemize}

\item routing functions:
	\begin{itemize}
	\item routes are fixed*: routes, for each agent type, are established before and maintained during the simulation runs. The sequence of markers/points that have to be reached is fixed;
	\item routes can dynamically change (by filtering or by verified conditions in the simulation): pedestrian preferred routes can dynamically change if some conditions inside the simulation are fulfilled (perception of alarms, excessive congestion or even more complicate combination of conditions defined by the user);
	\item pedestrian route choice to reach a specific target*: pedestrians can be distributed to different targets by percentage split, by less occupancy or shortest distance;
	\item possibility to filter pedestrians by pedestrian characteristics: pedestrians can be filtered by kind, destination, visited areas, actions they are performing (waiting, queuing);
	\item dynamic assignment/potential: the possibility of routing the passengers following a criteria that overrides shortest path and privileges quickest time, in relation to dynamic variations of conditions detected in the model.
	\end{itemize}
\end{itemize}

It should be possible to apply copying, pasting, scaling and translating to all the modelling and the routing objects as for the obstacle objects.

\subsubsection{OD-Matrix Input and Manipulation (subset 1\emph{(c)})}

As we have stated above, the possibility to import, manipulate and also export the demand (normally in the form of an OD-Matrix) is a very useful feature together with all simulation data and settings. Variation on the demand, also to represent different scenarios, are time-spending, and the possibility to intervene directly in the same environment and to export the variation in a format that is importable for future use (and the possibility to perform consistency checks on what has been put into the model) are time saving functions.

\begin{itemize}

\item OD-matrix input:
	\begin{itemize}
	\item OD-matrix input from worksheet file*: an OD-matrix (that specifies arrival profiles) can be elaborated and then imported in the simulator as .csv file or other formats;
	\item import data as timetable: arrival profiles can be imported as timetable (i.e. arrivals are injected at specific times);
	\item import as a spread over an interval: arrival profiles organized in specific time interval can be spread over customizable intervals of time (i.e. cumulative 5 minutes counting can be spread over 300 secs);
	\item supply types import: supply types can be defined in worksheet and then imported in the simulation. A supply type indicates, for each demand input, the percentage distribution of different pedestrian types that have to be produced for each entrance;
	\end{itemize}

\item OD-matrix manipulation: the platform allows editing the demands (adding, deleting, spreading, increasing, setting a specific frequency) directly in the platform without the necessity to import the new data from external source;

\item OD-matrix export into worksheet or other formats' file*: the OD-matrix or some arrival-departing profiles can be selected and exported into an worksheet format;
\item management of multiple settings: the OD-matrix imported in the simulation can be modified (for all input areas) by percentage and different settings can be stored in the system.
\end{itemize}

\subsubsection{Evacuation Studies}

Evacuation studies are very indicative of the performances of evacuation plans and can provide very useful insights to 
facility managers on total egress time and of the effectiveness of planned exit routes, predicting bottlenecks and criticalities of the structure (i.e. stairs).

\begin{itemize}
\item Evacuation mode: it is possible to trigger an \lq\lq evacuation mode\rq\rq\ in which all pedestrians, in spite of 
their specific targets, aim to the closest exit route;
\item it is possible to set a reaction time to the alarm*: evacuees can be set to take a fixed or variable amount of time before reacting to the evacuation alarm;
\item familiarity with the environment: it is possible to set a degree of knowledge of the environment for specific portions of evacuees 
population;
\item smoke data can be imported into the evacuation scenario: smoke data can be imported from other simulation software specifically 
dedicated to the study of fluid and gas dynamics.

\end{itemize}

\subsection{PEDESTRIANS: Requirements (subset 2)}

The majority of the platforms cited available on the market are based on agent paradigm \cite{russell1995artificial} in which the pedestrian dynamics is the result of micro-interactions between single individuals/pedestrians (agents) and the environment. Agents in agent-based simulation are clearly separated from the environment \cite{bandini2009agent} (differently from other approaches like cellular automata \cite{Schadschneider2002} or models more related to physics \cite{Helbing1995}). Agents can be of the reactive type (stimulus-response), cognitive/deliberative or hybrid \cite{wooldridge1995intelligent}. Generally, for computational reasons, in these platforms agents are mainly \lq\lq reactive\rq\rq, so they do not have representation of the environment and they do not have an \lq\lq inner schedule\rq\rq\ of their actions \cite{ferber1999multi}. It is the environment that, acting on them, maneuvers their actions through its elements. These agents, nevertheless, have some characteristics and they also can undertake independently some specific actions.

\begin{itemize}

\item Physical characteristics:
	\begin{itemize}
	\item speed is distributed: a \lq\lq preferred\rq\rq\ speed distribution is applied to each set of agents generated in the environment. The speed is assigned following a probabilistic distribution. Speed is customizable for specific types of agents;
	\item size is customizable: size can be varied and their space occupancy and speed distribution varies accordingly;
	\item luggage*: luggage of different dimensions can be assigned to each agent and their speed distribution varies accordingly;
	\item Persons with Restricted Mobility (PRMs): PRMs can be represented  (i.e. wheelchairs) in the model;
	\item agent libraries: a library of pre-defined agent types characterized by a specific speed distribution, variable size and other characteristics is already present between the simulation tools;
	\end{itemize}

\item actions:
	\begin{itemize}
	\item collision avoidance*: agents avoid collision with other agents and with obstacles present in the environment applying collision avoidance strategy. In the case of objects (i.e. walls) they do not walk along the wall or bump into objects (edge effect), but they avoid obstacle and other agents changing trajectory in advance;
	\item moving towards a target*: agents move by preferred/shortest path towards the assigned target;
	\end{itemize}

	\item planned activities: scheduled activities can be performed, in fixed or variable order, by the agents;
	\item knowledge of the environment: agents have knowledge or partial knowledge of the environment that allows them to take decisions and to perform some reasoning to select the best action to undertake to achieve their target. Pedestrians therefore, considering the available resources of the environment, maximize their utility in relation to the schedule tasks (i.e. if a long queue happens at the bottom of an escalator, on the basis of their knowledge they must know that they can take a longer but quickest route to their destination);

\item perceptive capabilities:
\begin{itemize}
\item perception of obstacles;
\item perception of other agent's presence;
\item perception of density;
\item perception of signals (visual or acoustic);
\end{itemize}

\item groups and social characteristics: possibility to set groups (families, friends) that tend to remain together are representable in the simulation.

\end{itemize}

\subsection{OUTPUT: Requirements (subset 3)}

\subsubsection{Analysis of Simulation Outputs}

In our experience we have noticed that in some simulation platforms more importance is attributed to the visualization of pedestrian dynamics than to the possibility of extracting analytical data and of analyzing outputs in depth. In the list below we have conceptually classified the kind of outputs in relation to the measure we refer to:

\begin{itemize}

\item density related:
	\begin{itemize}
	\item local density*: pedestrians per square meter in a specific portion of the environment;
	\item Level of Service (LOS)/ Cumulative Mean Density (CMD)* : Fruin's Level \cite{fruin1992designing} of service, HCM LOS \footnote{Highway Capacity Manual US standard} and other in a specific portion of the environment, also in the form of CMD that is the relative average density measured across a specific interval of time around a specific portion of space (or single cell if space is discretized) \cite{castle2011comparison};
	\item utilization of space*: related to the number of time that a location has been occupied in a specific period of time;
	\end{itemize}

\item time-related:
	\begin{itemize}
	\item total, average and individual transfer times between two lines in a specific time interval*: possibility to measure transfer times 
	between two or more poly-lines;
	\item total, average and individual transfer times inside a specific area in a specific time interval*: possibility to measure transfer times inside a specific area;
	\item total, average and individual queuing time in a specific area in a specific time interval*: possibility to count people performing a 
	specific action (see also the filtering in previous Sections);
	\end{itemize}

\item time-density related: Service Factors (SF) is a sort of weighted LOS that takes into consideration not only the LOS experienced by the user, but also the percentage of user that have experienced each level of density and for how much time; 

\item distance-related: total and average distance covered by pedestrians*. Inside a specific area, total distance and average distance covered by pedestrians are calculated with the possibility to filter the measurement by type, action and so on;

\item flow and counting-related:
	\begin{itemize}
	\item counting across two lines (or more)*: counting the number of pedestrians/flow rate that cross two (or more) poly-lines;
	\item counting across a single line*: counting the number of pedestrians/flow rate that cross one single line or poly-line;
	\item counting inside an area*: counting the number of pedestrians inside a specified area;
	\end{itemize}

\item analysis filtering*: possibility to filter by kind of agents, by a user defined selection of agents, by destination, by visited destination, by action performed (waiting, queuing, etc.);

\item performance-related: Social Costs analysis associates an economic value to the  time spent by passengers on the basis of the activities they are performing. Different activities weighted differently in relation to their desirability (e.g. waiting and queuing time are weighted more than walking time in a train station scenario). The social cost for a specific scheme sums up all the costs calculated for each passenger in the peak time;

\item analysis and OD-matrix can be exported into worksheet*: this utility is especially useful for future re-import of the matrix into the model but also to perform consistency check of the imported demand (if the routing is dynamics in the model);

\item automatic generation of analysis: it is possible to set the software to perform customized analysis at specific intervals.

\end{itemize}

\subsubsection{Presentation Tools}

Presentation tools support the communication and visualization of results and elaborated analysis in an easy and understandable manner, also to non-expert users.

\begin{itemize}

\item Video output in .avi/.mov or other formats*: it is possible to generate video output at specific simulation times and basic video editing is supported;

\item screenshots in .jpg/.bmp or other formats*: it is possible to export screenshots in different formats;

\item maps:
	\begin{itemize}
	\item density maps*: the platform allows to generate density maps and the associated legend;
	\item SF maps: the platform allows to generate SF maps and the associated legend;
	\item utilization of space maps*: the platform allows to generate utilization of space maps and the associated legend;
	\item time maps*: the platform allows to generate time maps where different colours highlight when a position in the simulated
	environment was last occupied. These maps are particularly informative in evacuation studies as they allow to mark with a specific colour the areas of the environment that where still occupied after the maximum expected evacuation time;
\end{itemize}

\item chart generation is supported: analysis can be visualized in chart format directly in the platform before the export into worksheet;

\item definition of a color scheme to visualize/track specific entities*: it is possible to highlight with specific colors precise entities by filtering by kind of pedestrians, actions, final destination;

\item visualization of paths of tracked individuals*: possibility to visualize the trial of tracked individuals;

\item time stamp visualization*: it is possible to visualize the simulation time and to impress the time stamp on videos and screenshots 
before exporting;

\item automatic generation of maps and charts: it is possible to set the software to perform customized maps and charts at specific intervals.

\end{itemize}

\subsection{Additional Features}

Beyond all the aspects that we have analyzed above, we want to discuss further characteristics that, although not directly involved in the modelling and analysis process, constitute important aspects that require an evaluation. These aspects are related to the interaction between vehicle and pedestrians, the robustness and the technical aspects of the platform and the validation of the displayed dynamics.

\subsubsection{Vehicle and Pedestrian Interaction}

Many simulation platforms propose pedestrian simulation module as an addition to a vehicular traffic module (that often constitute their main market). Especially in studies related to street surfaces the possibility to give an account of vehicular and pedestrian interaction is very important, in particular in presence of zebra crossings (without traffic lights) and in case of non compliant road users (whose behavior, that might change in relation to geographical habits, definitely affects vehicular traffic queue and jams formations in peak hours). 

\begin{itemize}

\item Vehicular traffic (cars, vans, buses and so on):
\begin{itemize}
\item simulated in the software: it is possible to simulate interaction between vehicles and pedestrians in the same simulation 
platform (e.g. some software offer a very good integration of pedestrian and vehicular traffic);
\item simulated in a different software but with import output: it is possible to simulate vehicular traffic in another environment and 
then to import the results and to combine them with pedestrian traffic outputs;
\end{itemize}

\item bicycle traffic: it is possible to simulate bicycle traffic, customized in the platform;

\item interaction between vehicles and pedestrians: car and vehicles perform collision avoidance in shared spaces or in un-signalized crossings;
\item non compliant road users, with the possibility to define a percentage of undisciplined pedestrians: it is possible to define a percentage of pedestrians that does not respects traffic signals (i.e. traffic lights) and therefore the consequent impact on vehicular traffic.
\end{itemize}

\subsubsection{Robustness and Technical Details}

This part is still under development, we would consider a pedestrian software robust if it can handle a high number of agents in 
co-presence, and if it can run in a reasonable amount of time, without frequent crashing. In studies that involve very large number of people this aspects can represent a limitation that might also compromise the possibility to bring a study to its end. Also, in this Section, we would like to introduce some technical details related to the phase of simulation runs:

\begin{itemize}
\item automatic consistency check of the model and support to the debugging process;
\item number of maximum agents handled simultaneously in the simulation: the platform can handle at least 100.000 agents at the same in the peak hour;
\item software crashes during specific operations and running times;
\item general usability of the interface; 
\item multiple runs: possibility to run parallel simulations;
\item simulation seed: possibility to change (manually or automatically) the simulation seed to perform comparative studies;
\item batch runs: possibility to launch batch runs.
\end{itemize}

\subsubsection{Validation}

The process of validation of the platform can be done in different ways:

\begin{itemize}
\item simulation platform is validated against real observations: a robust validation campaign has been performed to guarantee the 
validation of the platform in all possible general situations;
\item simulation default settings reflects Fruin's LOS \cite{fruin1992designing} for all situations: simulation calibration has been performed for paradigmatic cases (i.e. one or two way flow in a corridor) to reflect what predicted (in terms of LOS) by standards (i.e. Highway Capacity Manual or different \cite{rouphail1998capacity});
\item simulation requires case by case calibration*: simulation requires calibration case by case but clear guidelines are provided;
\item the software validation is certified: an external third party has certified the validation of the software.
\end{itemize}

In relation to the validation of a simulation platform also the corroboration of the validation data by an external agency or institution with no conflict of interest with the producer of the software is considered of value. From this point of view, the technical note NIST 1822\footnote{\url{http://www.nist.gov/customcf/get_pdf.cfm?pub_id=913642}} proposes to initiate a methodological debate inside the discipline of simulation practice with the aim of defining more rigorously the verification and validation standards applied to models, although the NIST 1822 for the time-being refers specifically to evacuation dynamics.

\section{Interpretation of the Completeness of the Checklist in an Automatized Worksheet}

To provide an easier usability of the checklist described above we have automatized the process of evaluation in a worksheet. 
A tab has been created for each category of evaluation (i.e. \lq\lq 2D Environment\rq\rq, \lq\lq Modelling and Routing\rq\rq, etc.) and for each element of the list it is possible to assign three different values (from a drop down list): \lq\lq yes\rq\rq, \lq\lq no \rq\rq, \lq\lq under development\rq\rq. The checklist includes several items, in our opinion all of them are useful features to perform complete and robust studies that cover the major aspects of pedestrian studies in all possible scenarios. Nevertheless a subset of the list constitutes what we think that represents the minimum sufficient requirements according to market's needs. 
The selection of these last voices (signed with * in the list) might be open to discussion. For this reason we have left, in our worksheet, the possibility to customize this selection. On the basis of these considerations we propose the method that we apply internally and that aims at providing two kinds of feedback: the first aims at checking the fulfillment of the minimum requirements we have designed, and the second instead measures the completeness considering the whole of the possible functions that we have listed. 
In relation to the first analysis, for each category the lack of one single mandatory requirement will be enough to label the category as insufficient. The second method, instead, measures the percentage of the completeness (in relation to the whole of listed items). 

A beta demo of the worksheet, filled in with sample data is available at: \url{https://www.dropbox.com/s/igf1f9foy9z8h0z/ped_chk.zip}. Any comments or feedback related to this work is welcome.

\section{Conclusions and Future Works}

In this contribution we have proposed a checklist for the evaluation of pedestrian modelling software. In recent years many commercial platforms have been launched on the market and we believe that a guidance for the evaluation of these products could help the final users in the choice of the appropriate software package but also the software's developers at moving towards a standardization of functions.

This paper aims at representing a point of view in an open discussion on this topic and at encouraging contributions for further refinement of the selected criteria. Future works include a more detailed examination of the software usability, together with a more engineered procedure to evaluate the robustness of these products. We would like also to cover aspects related to the evaluation of platform costs, licensing schemes and quality and accessibility of customer assistance. Moreover we would like to extend the checklist to include also similar detailed evaluation items for vehicular simulation software. 

Finally, we would like to collect opinions in relation to our evaluation criteria. Nevertheless we consider the checklist sufficiently complete to be already used as a valid support for practitioners in the process of evaluation of pedestrian dynamics simulation software.

\bibliographystyle{splncs}
\bibliography{TRBLaTeX}

\end{document}